# Impurity in low energy Ar$^+$ ion beam is the cause of pattern formation on Si

**Dipak Bhowmik[1], Manabendra Mukherjee[2] and Prasanta Karmakar[1]***

[1]Variable Energy Cyclotron Centre, HBNI, 1/AF, Bidhannagar, Kolkata -700064, India

[2]Saha Institute of Nuclear Physics, HBNI, 1/AF, Bidhannagar, Kolkata -700064, India

We report the decisive role of reactive ion impurities in low energy Ar$^+$ ion beam on surface nanopattern formation. The source of experimental inconsistency in pattern formation by low energy (few keV to 10's of KeV) Ar$^+$ ion beam has been identified by irradiating Si surface at an oblique angle with pure and impure Ar$^+$ ion beam of energy 3-10 keV. No well-defined patterns are observed for mass selected pure Ar$^+$ ion bombardment, whereas well defined periodic ripple pattern is formed by the same experimental condition with impure mass unanalyzed Ar$^+$ ion irradiation. The contaminants in mass unanalyzed beam specifically reactive nitrogen, oxygen and carbon play the main role of pattern formation by introducing chemical instability on the Si surface. The surface morphology of the irradiated Si surfaces is examined by Atomic Force Microscopy (AFM). The surface contamination and corresponding chemical compound formation are investigated by X-ray photoelectron spectroscopy (XPS).

Keywords: Si, ion beam, ripple-patterning, AFM, XPS.

*Corresponding author e-mail: prasantak@vecc.gov.in

**Introduction**

The pattern formation on Si surface by energetic ion bombardment has brought great interest in scientific research for its potential applications. Several experimental observations of ripple pattern formation have been reported on Si surface by varying ion species, energy (250 eV - 1 keV) and incidence angles [1-9]. As the Ar ion beam is very common for surface cleaning and depth profiling in surface science experiments, several groups have reported pattern formation on Si surface by very low energy $Ar^+$ ion (500 eV to 2 keV) bombardment [2,4-14], whereas the study is limited in the medium energy range 3 – 100 keV [15-29] as well pattern is observed only for $Ar^+$ ion energy greater than 20 keV [20-29]. It is interesting that nanopatterning on Si by Ar beam is suppressed in the energy range 3-20 keV [15-19,30].

For very low energy (100 eV to 2 keV) $Ar^+$ ion beam, ion guns are generally attached to the vacuum chambers without the mass filtering system, whereas in typical ion implanter the energy is above 3 keV, and the beam is usually mass analyzed and isotopically pure. The unfiltered ion beam may contain N, C, O and other common impurities. Pan et al. [31] observed silicon carbide (SiC) formation in an ultra-high vacuum chamber by low energy (1.5 keV) $Ar^+$ ion incorporation; although they did not study the pattern formation and role of contamination induced chemical effects on the pattern formation. Ziberi et al. also admitted that the dot pattern formation by noble ion bombardment might be due to the unintentional contaminations from the various sources [2,10,32,33]. Hofsass and Engler pointed out the effect of contaminations on pattern formation, though that contamination comes from the target or unintentional incorporation of impurity on the sample during ion bombardment [34-36]. Nevertheless, the contaminations in ion beam and consequent effects on nanopatterning have been overlooked.

In this article, we have investigated the pattern formation on Si by both the mass analyzed as well as unanalyzed ion beam and explored why the Si surface patterns are found in low energy unanalyzed beam and suppressed for 3-10 keV mass analyzed $Ar^+$ ion beam. We report that the reactive impurities in the ion beam play the vital role in the pattern formation.

## Experimental

The commercially available Si (100) wafers of size 1 cm × 1 cm after cleaning with trichloroethylene in an ultra-sonic cleaner were irradiated with ion energy $3 - 10$ keV $Ar^+$ ion beam at oblique angle incidence from a 2.4 GHz ECR ion source of the Radioactive Ion Beam Facility at Variable Energy Cyclotron Centre (VECC), Kolkata. The irradiation was carried out in two different ways one before the dipole magnet, i.e. by mass unanalyzed ion beam and another after the dipole magnet, i.e. by mass analyzed ion beam. The schematic diagram of the irradiation experimental set up is depicted in figure 1. The beam was collimated to get a uniform beam of 8 mm diameter on the sample surface which was maintained with proper secondary electron suppression for both the system. The pressure in the ECR chamber was around $5 \times 10^{-7}$ mbar and in the target chamber, it was $3 \times 10^{-7}$ mbar during the ion beam experiment. The morphology of all the irradiated samples was investigated in air using Bruker Atomic Force Microscopy (AFM), Multi-Mode V at VECC, Kolkata.

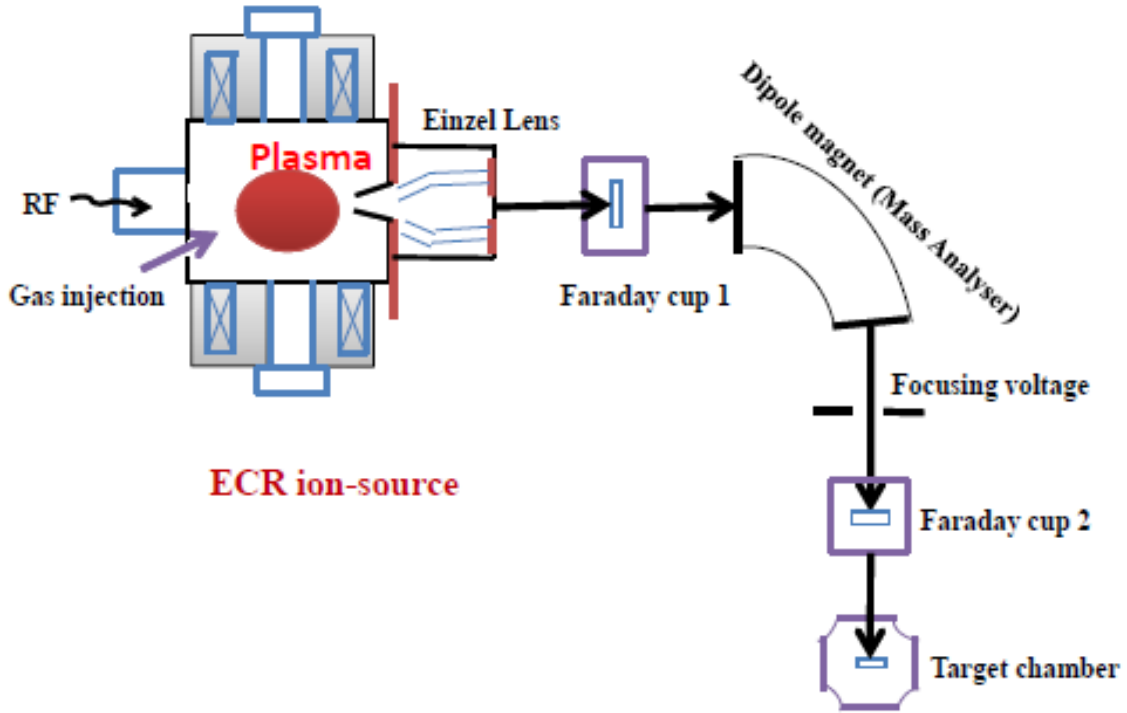

**Figure 1** : Schematic diagram of the ECR ion source and low energy separator.

The compositional change of the irradiated Si (100) samples with respect to virgin Si surface was investigated by X-ray photoelectron spectroscopy (XPS) using an Omicron Multi-probe (Omicron Nano Technology, UK) ultrahigh vacuum(UHV) system (base pressure $\sim 5.0 \times 10^{-10}$ mbar). The system is equipped with a monochromatic Al $K_\alpha$ source with photons of energy 1486.6 eV for XPS measurement.

**Results and discussions**

Figure 2 shows the surface morphology of mass analyzed Ar$^+$ ion-bombarded Si surfaces with different ion energy. Fig. 2 (a) – (d) show the AFM images of 3 – 10 keV Ar$^+$ ions bombarded Si surfaces with ion fluence 7×10$^{17}$ ions/cm$^2$ at an incidence angle 60$^o$ with the surface normal. The irradiated surfaces get amorphized after the irradiation and do not show any pattern, but a rough surface with rms roughness below 1 nm. The rms roughness of the ion bombarded surfaces increases slightly with ion energy which is shown in Figure 1 (e). The absence of pattern on Si surface by low energy (keV) $Ar^+$ ion bombardment at oblique angle was also reported earlier [19,30]. But, the explanation for the absence of pattern formation in these energy regimes is not substantial as the patterns are easily observed by other groups at lower energies (< 2 keV) [1-9]. Ziberi et al. reported the ripple pattern formation on Si by noble-gas ion beam ($Ar^+$, $Kr^+$, $Xe^+$) below 2 keV ion energy and observed the absence of pattern by $Ne^+$ ion beam [37]. They concluded that the mass of ion plays a vital role in pattern formation. If mass, energy, angle of incidence and fluence of the projectile are only the factor of pattern formation, then at intermediate energy (~10 keV) Ar should form the pattern. This inconsistency is still unexplained for low energy $Ar^+$ ion beam induced pattern formation. It is observed that the kinematics of the $Ar^+$ ion with Si surface is not very much different for ion energy 1.3 to 10 keV, and indicate the possibility of ripple formation in all this energy [30]. So, the absence of pattern formation in the medium ion energy regime indicated that only the kinematics is not sufficient to develop well-defined patterns on the surface, hence, the additional sources of instability, i.e., initial perturbation[38], presence of surfactant [34] or beam induced impurity [19,39] are to be considered. To examine the source of instability, the chemistry of $Ar^+$ ion-bombarded Si surface is investigated by X-ray Photoelectron Spectroscopy (XPS).

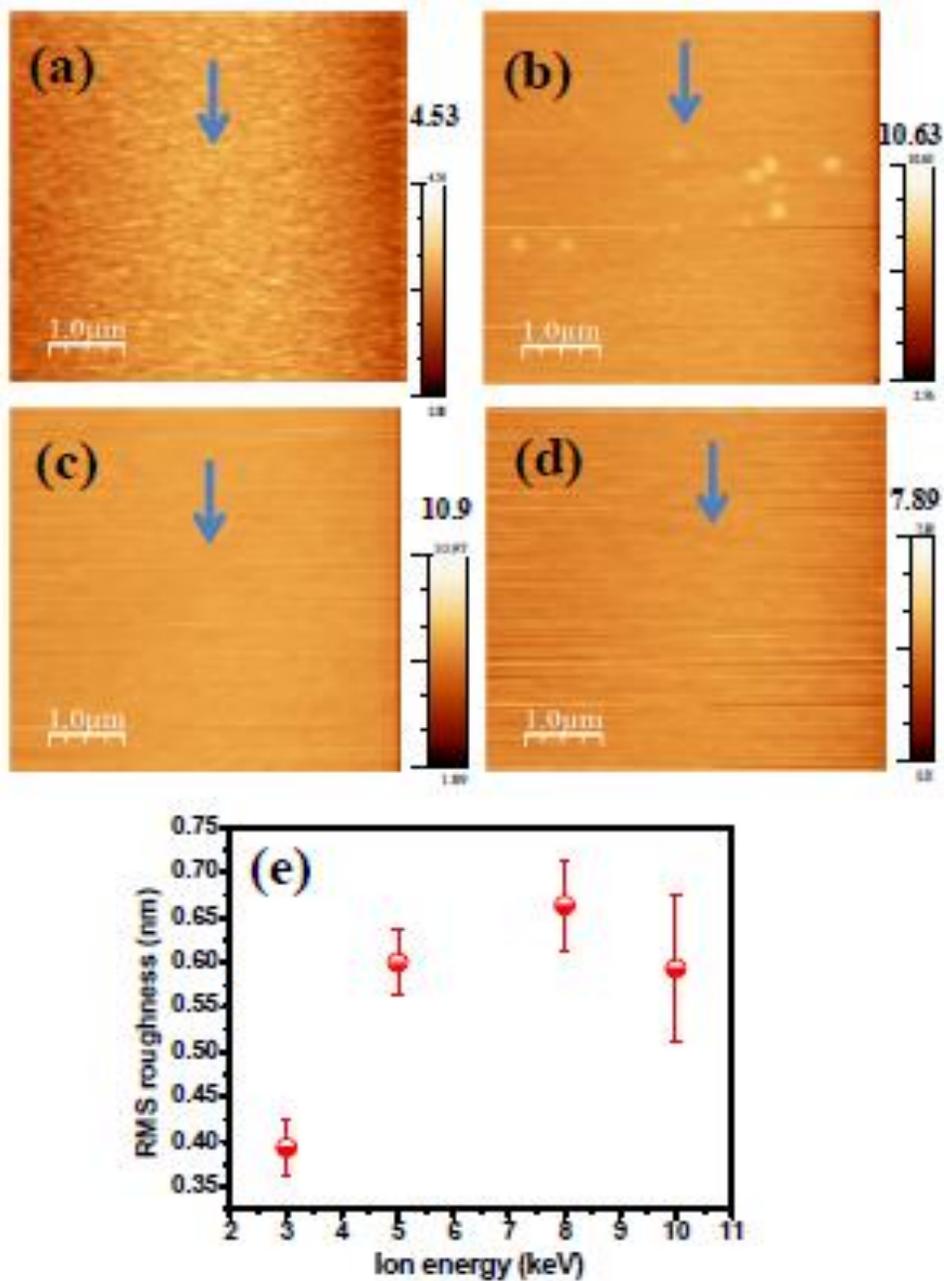

**Figure 2** : AFM images of mass analyzed Ar bombarded Si surface with ion energy (a) 3 keV, (b) 5 keV, (c) 8 keV and (d) 10 keV at oblique angle incidence 60˚ with constant ion fluence $7 \times 10^{17}$ ions/cm$^2$. The arrow in each image indicates the ion beam direction and Z scale is shown with each image. (e) The RMS roughness of mass analyzed Ar bombarded Si surfaces with ion energy.

Figure 3 shows the XPS spectra of mass analyzed 5 keV and 10 keV $Ar^+$ ion-bombarded Si as well as virgin Si surface. It is clear from the figure that the surface chemical state of Si has not changed due to 5 or 10 keV mass analyzed $Ar^+$ ion bombardment. The oxide peak for all three surfaces in figure 3 is for the native oxide layer, formed due to air exposure during sample transfer from the implantation chamber to the XPS system. The binding energy and concentration of elemental Si and $SiO_x$ for all the surfaces are mentioned in figure 3. Similar oxide state at this binding energy was observed for oxide growth in Si NWs [40]. The implantation of Ar atoms is confirmed by high-resolution Ar 2p core level XPS spectrum (figure 4 a). Ar $2p_{1/2}$ and $2p_{3/2}$ are clearly resolved with separation 2.1 eV which indicates inertness of Ar within the Si matrix. The distribution of implanted 5 keV  Ar atoms  at $60^o$ in Si as a function of depth  is simulated by TRIM [41] and presented in figure 4 (b). Ar atoms during irradiation transfer its energy to the target material and penetrate up to a certain depth. Because of its inertness, it is only trapped in the Si surface. The two trapping mechanism for Ar in Si was given [42]. In the first mechanism, the Ar atoms are trapped in vacancy sites whereas in the second mechanism it sits in interstitial sites. At low energy, Ar atoms are trapped mainly in interstitial sites. Similar Ar ion trapping was observed in Si for low energy ion bombardment [31]. Bradley and Hofsaas [43]  predicted that the presence of implanted species also responsible for surface instability, however, the impact is lower than the sputtering and mass redistribution effects. Ripple formation on Si surface by Ar[+] ion beam is only possible if the surface instability exceeds a threshold level. The threshold level can be achieved by Ar[+] ion energy more than 10 keV [20-27 28,29], or at very high fluence[16,17]. Therefore, Si ripple formation by Ar[+] ions of energy up to 10 keV at moderate fluence is unlikely as sputtering, mass redistribution of Si target atoms and presence of inert implantations are not sufficient to generate instability for surface pattern formation. Hofsaas el al. also showed that due to very low growth rate of parallel ripple pattern, it is unlikely to observe the pattern by Ar ion irradiation between 1.3 -10 keV for typical ion fluence up to about $10^{18}$ ions/ cm$^2$ [30]. Presence of additional reactive atoms/ ions and subsequent surface chemical change may trigger the instability by altering the sputtering and mass redistribution effects, which enhances the growth rate of the parallel ripple patterns.

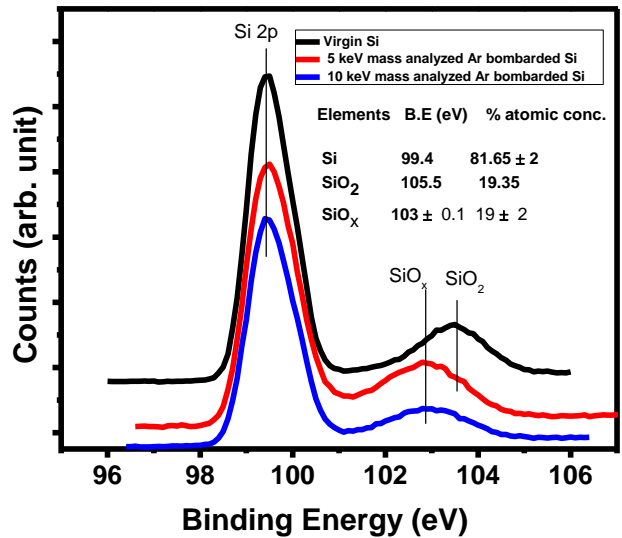

**Figure 3** : High resolution Si 2p core level spectra of  virgin Si, 5 keV and 10 keV mass analyzed Ar bombarded Si.

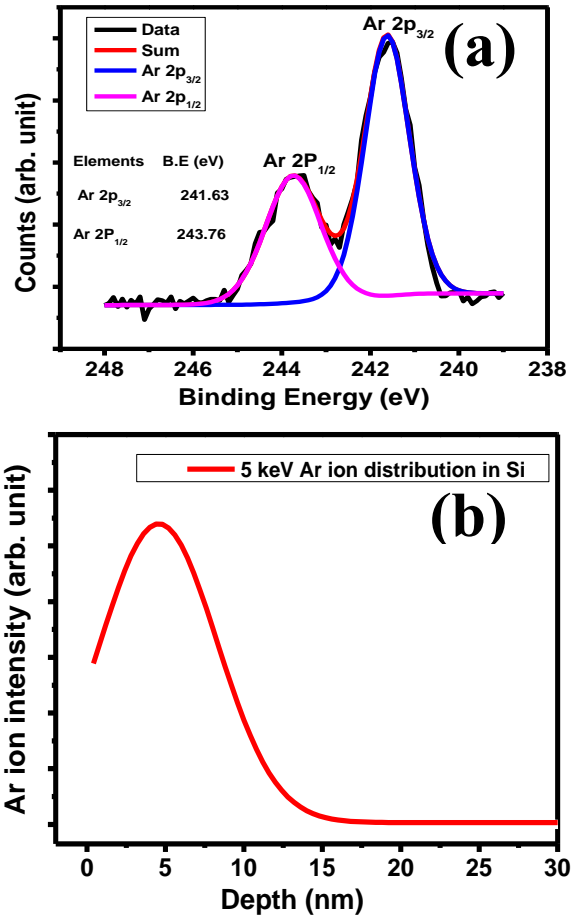

**Figure 4**: (a) High resolution Ar 2p core level spectra of 5 keV mass analyzed Ar bombarded Si surfaces. (b) TRIM calculation of implanted Ar distribution for 5 keV Ar bombarded Si at 60[o].

To verify our assumption, we bombard the Si surfaces by unfiltered impure Ar$^+$ beam similar to ion bombardment with ion guns without mass filtration. We kept the same ion beam parameters and ion surface geometry as before. The experiments were performed before the dipole magnet (mass analyzer) as shown in Figure 1. The AFM morphologies of Si surfaces after the bombardment with unanalyzed 3-10 keV Ar$^+$ beam is shown in figure 5 (a) – (d). All the bombarded surfaces show well periodic nanoripple pattern. The rms roughness and ripple wavelength of the bombarded surfaces with ion energy are shown in figure 5 (e) and (f). The rms roughness of the surfaces with ion energy changes within ± 1 nm, whereas the ripple wavelength increases with ion energy as usual.

To investigate the possible reason of nanopattern formation with unanalyzed ion beam, we took the mass spectrum of the ion beam as well as investigated the irradiated Si surface by XPS. The XPS survey of 5 keV and 10 keV Ar$^+$ ion bombarded as well as virgin Si surfaces with and without mass analyzer is shown in figure 6. The virgin Si surface contains Si, oxygen (O1s) and carbon (C1s). The presence of carbon and oxygen in Si is common [31]. Oxygen comes from the native oxide layer if it is exposed to air and a trace amount of C remains during the Si wafer processing. We compare the XPS survey spectrum of virgin Si, and the Si bombarded with mass filtered and unfiltered 5 & 10 keV Ar ions. For mass filtered Ar$^+$ ion bombardment, Ar peak is observed in addition to Si, C and O peaks. It shows that the surface contains the same elements as virgin Si with additional Ar peak. But, in the case of unanalyzed Ar bombardment, a new N peak, as well as the increased intensity of C and O, is found. It proves the presence of contamination in mass unanalyzed ion-bombarded Si surface.

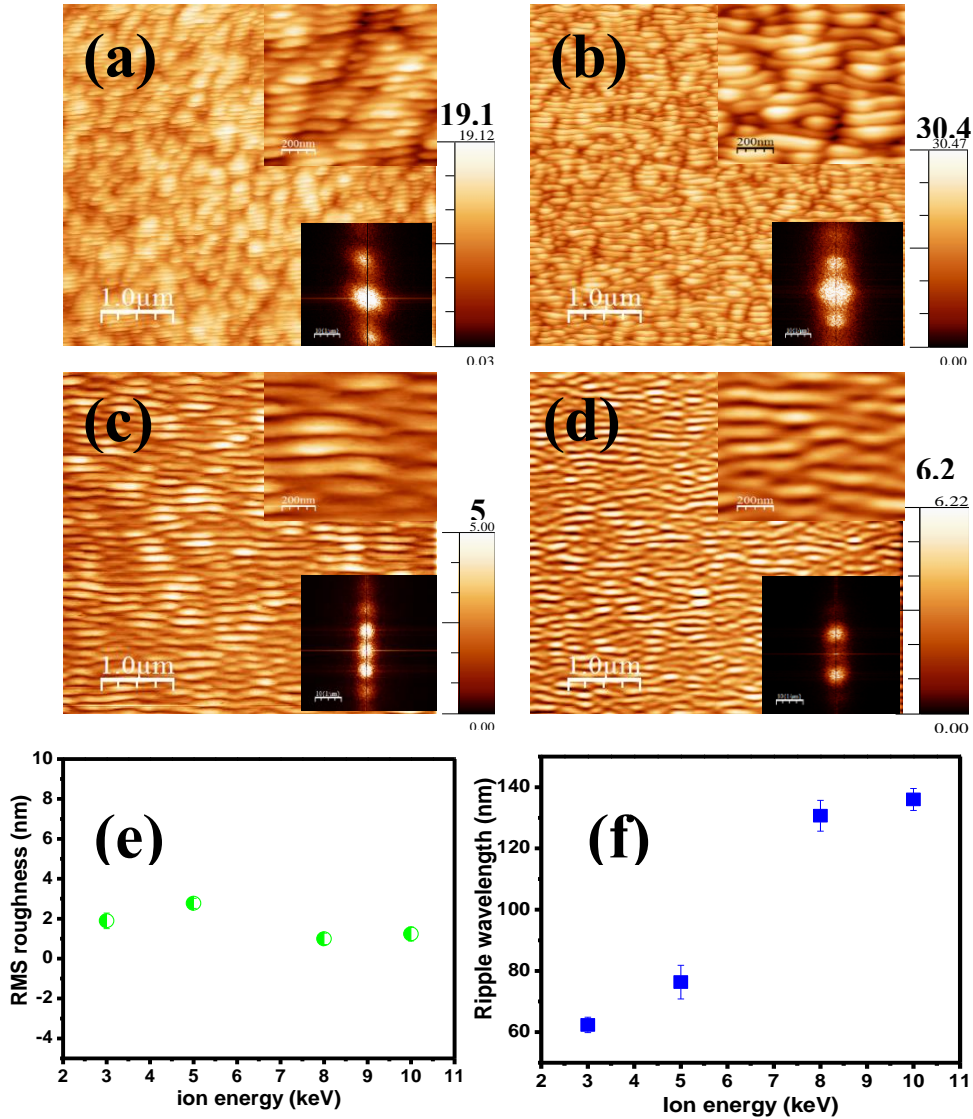

**Figure 5** : AFM images ($5\mu m \times 5\mu m$ ) of without mass analyzed Ar bombarded Si surface with ion energy (a) 3 keV, (b) 5 keV, (c) 8 keV and (d) 10 keV at oblique angle incidence 60° with constant ion fluence $7\times10^{17}$ ions/cm². The FFTs are shown in the corner of each image showing parallel mode ripple pattern formation. The $1\mu m \times 1\mu m$ scan AFM images are also shown in the upper corner of each AFM images. The arrows indicate the ion beam

direction and also Z scale is shown near each image. (e) The surface RMS roughness and (f) ripple wavelength of mass unanalyzed Ar bombarded Si surfaces with ion energy.

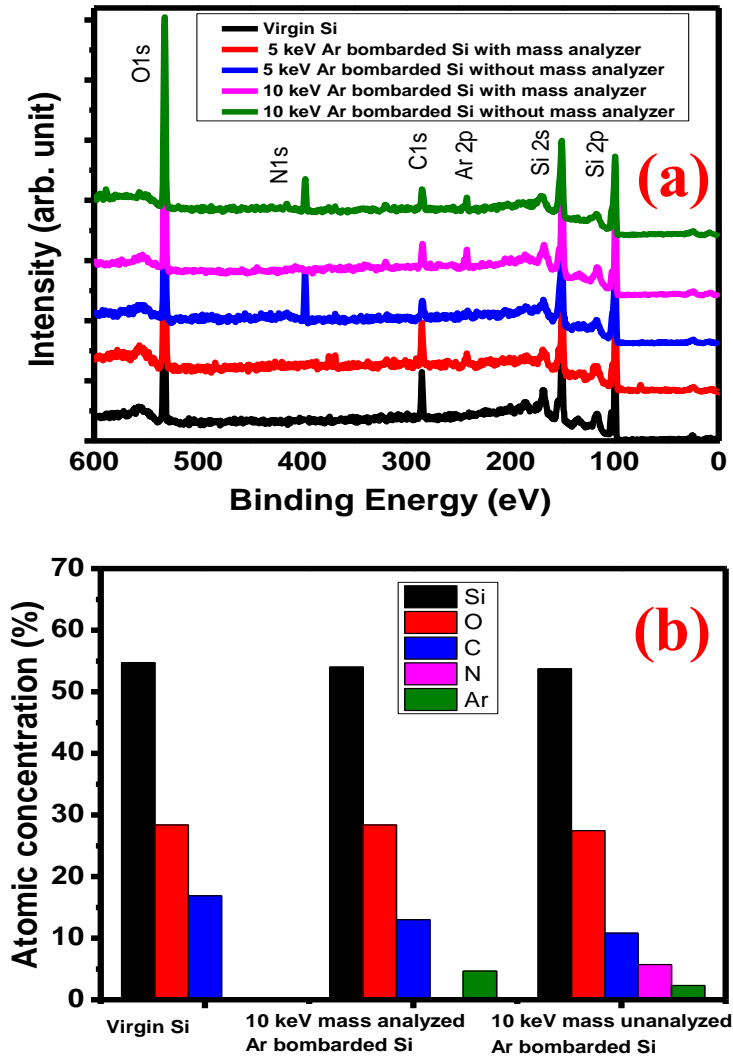

**Figure 6** : (a) XPS survey of virgin Si and ion bombarded Si surfaces with mass analyzed beam and without mass analyzed beam. (b) Atomic concentration of different elements in virgin Si, 10 keV mass analyzed and mass unanalyzed Ar bombarded Si surfaces calculated from XPS survey.

To identify and quantify the presence of contamination in the unfiltered beam, we recorded the mass spectrum of the ion beam extracted at 10 keV by the same analyzing magnet. Although the ion source was filled with pure Ar gas, we find other common species in the form of ions. The mass spectrum for 10 keV beam is shown in figure 7. The spectrum shows H, N, O and C ions along with the Ar ion. So, the unanalyzed Ar ion beam that bombarded the Si surface is a mixture of Ar, O, N, C and H ions. Therefore, all of these ions bombard the Si surface when the beam is not filtered by the analyzing magnet. This type of beam contamination is very common for almost all type of ion sources [44-46]. It is found that the mass spectra are consistent with the XPS survey data considering the sensitivity factor of the respective elements (figure 6b). The contaminations in the $Ar^+$ beam specifically C, O, N like reactive species change the chemical nature of the surface which generates additional surface instability during ion bombardment.

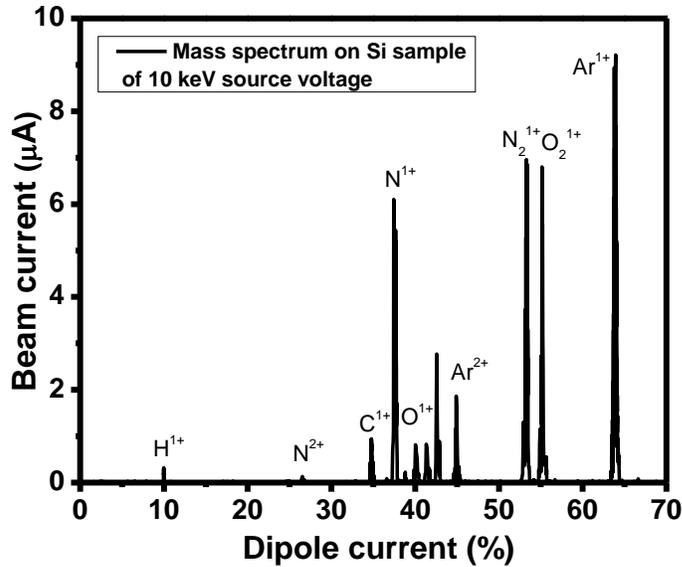

**Figure 7**: Mass spectrum on Si (100) sample by 10 keV source voltage extracted from ECR ion source.

The change of surface chemistry due to contaminated Ar$^+$ ion bombardment is further investigated in details by high-resolution XPS measurements. Figure 8 (a) and (b) show the Si 2p core level spectra of Si surfaces bombarded with unfiltered Ar$^+$ ion (5 & 10 keV). The Si 2p peak can be fitted by four p-type Gauss-Lorentz peaks (2p$_{3/2}$ + 2p$_{1/2}$) which contain elemental Si (B.E. = 99.2 ± 0.15 eV ), SiC (B.E = 100 ± 0.1 eV), Si$_3$N$_4$ (B.E. = 101.4 ± 0.2 eV) and SiO$_2$ (B.E. = 103 ± 0.2 eV). Also, the % area of unreacted Si and its compound are calculated from figure 8 (a) & (b) which are shown in table 1.

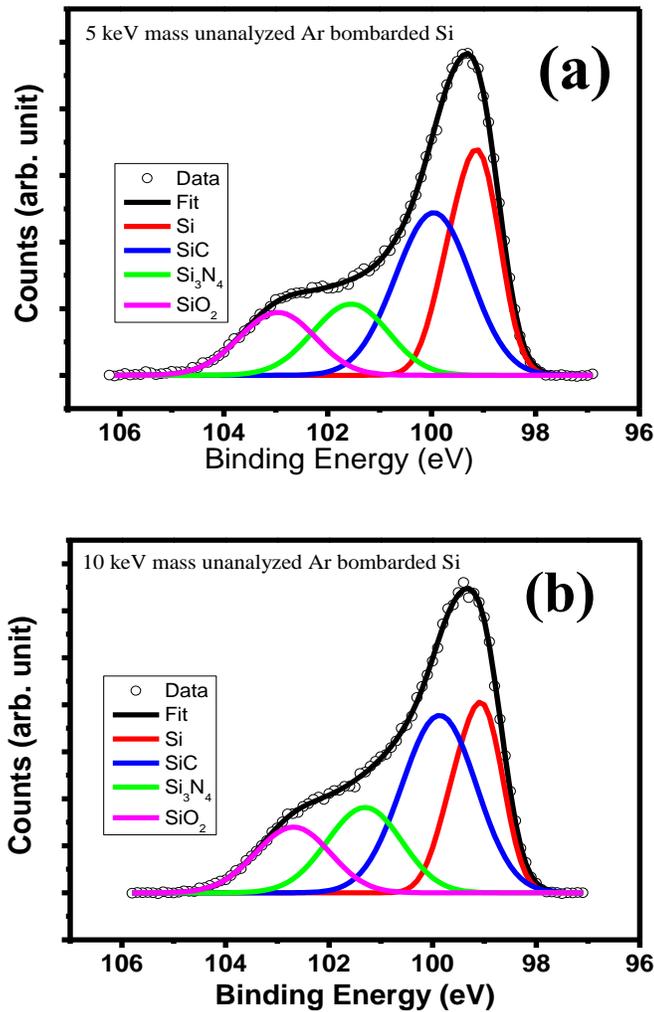

**Figure 8** : High resolution Si 2p core level spectra of (a) 5 keV mass unanalyzed and (b)10 keV mass unanalyzed Ar bombarded surfaces showing chemical compound formation.

As C is present in all the samples, we have taken high-resolution spectra for C 1s for virgin, and Ar bombarded Si surfaces. It is found that C beam plays a different role when implanted in Si compared to the carbon commonly present in Si as contamination. Figure 9 (a) and (b) show the core level C 1s spectra of virgin Si and 10 keV mass selected Ar bombarded Si surfaces. No chemical change of as presented C in Si is observed here. However, C 1s high-resolution spectrum from Si surface bombarded with unanalyzed contaminated (contains C ion also) beam shows chemical alteration of C 1s spectra. The spectrum shown in figure 9 (c) is fitted by three peaks, at 282.88 eV, 284.7 eV and 286.5 eV corresponding to silicon carbide (SiC), elemental C and hydroxyl (C-OH) respectively. The peak at 282.88 eV confirms again the SiC formation by $C^+$ impurity ions present in the mass unanalyzed ion beam. . The formation of $Si_3N_4$ and SiC during N and C ion bombardment to Si surface was also previously observed [19,39,47]. The table 2 shows the binding energies (B.E) of Si, SiC, $Si_3N_4$ and $SiO_2$ reported earlier and measured in the present study. The hump at 286.5 eV is due to hydroxyl adsoption which was also observed previously around this binding energy [48,49]. Similar hydroxyl (C-OH) hump is also present at slight higher binding energy for virgin and 10 keV mass analyzed Ar bombarded Si surfaces as shown in figure 9 (a) and (b). The high-resolution Ar 2p core-level spectrum for 5 keV mass unanalyzed Si surface is also shown in figure 9 (d) which displays the fine splitting of Ar $2p_{3/2}$ and $2p_{1/2}$ with binding energy separation 2.1 eV. It again shows inertness with Si as is observed before for Si surface bombarded with mass selected Ar ions (fig. 4a).

Table1. % area of unreacted Si and its compounds from figure 8 (a) & (b).

|  | Unreacted Si | SiC | Si₃N₄ | SiO₂ |
|---|---|---|---|---|
| 5 keV ion energy | 34 | 36 | 16 | 14 |
| 10 keV ion energy | 28 | 39 | 19 | 14 |

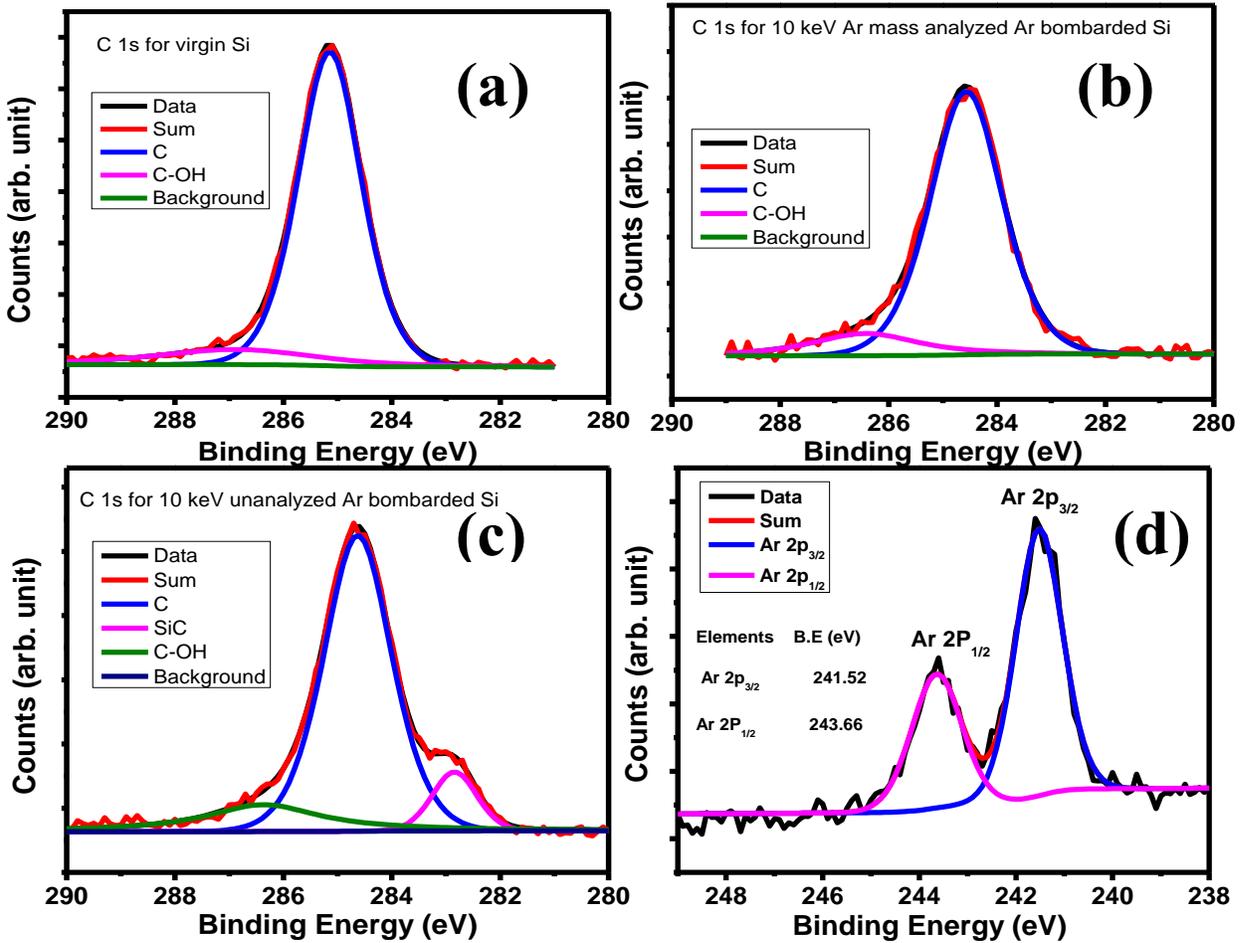

**Figure 9**: High resolution C 1s core level spectra of (a) virgin, (b) 10 keV mass analyzed and (c)10 keV mass unanalyzed Ar bombarded surfaces showing SiC formation. (d) Ar 2p core level spectrum for 5 keV mass unanalyzed Ar bombarded Si surface.

Table 2. Some reported binding energies (eV) of Si 2p, C 1s and their compound along with the present experimental value.

| Si region | | | | C region | | Ref. |
|---|---|---|---|---|---|---|
| Si | SiC | Si$_3$N$_4$ | SiO$_2$ | SiC | C | |
| 99.8 | 100.3 | | | 283.3 | 285.4 | [50] |
| | | 101.3 | 102.4 | | | [51] |
| 99.3 | 100.1 | | | 283.2 | 284.8 | [31] |
| 99.4 | | | 103.2 | | | [52] |
| | | 101.8 | 103.7 | | | [53] |
| 99.2 ± 0.15 | 100.0 ± 0.1 | 101.4 ± 0.2 | 103.0 ± 0.2 | 282.9 | 284.6 | Present work |

The absence of well-defined pattern by mass selected Ar$^+$ beam indicates that when pure kinematics induced instabilities are not sufficient, reactive contaminants in the unanalyzed primary beam introduce surface chemical inhomogeneity to generate required instability for pattern formation. Researchers have tried to explain the ripple pattern formation on the basis of curvature depended coefficients of height equation considering the effect of sputtering, mass redistribution and also the presence of implant species [5,30,43,54-58]. But, these theoretical improvements could not able to explain the absence of ripple pattern formation on Si by Ar$^+$ ion bombardment in ion energy (3-10 keV) [15-19] whereas it is easily observed in the lower energy (< 3 keV) with Ar$^+$ ion beam [2,4-14]. Although Hofsass [30] explained theoretically the absence of pattern formation, but still now no such experimental explanation has been reported. We have shown here that the presence of implant species and specifically the reaction of reactive implants with Si and consequent alteration of sputtering and mass redistribution of the chemically inhomogeneous surface might explain the paradox.

The height equation for ion sputtered surface is written as [59]

$$\frac{1}{J}\frac{\partial h}{\partial t} = C_{11}\frac{\partial^2 h}{\partial x^2} + C_{22}\frac{\partial^2 h}{\partial y^2} - \kappa \nabla^4 h$$

Here, $C_{ii}^{tot} = C_{ii}^{\ sputt} + C_{ii}^{\ C\,V} + C_{ii}^{\ impl}, \quad i = 1, 2.$

Hofsaas et al. [30] calculated the $C_{11}^{tot}$, considering the curvature dependent sputtering ($C_{ii}^{\ sputt}$), mass redistribution($C_{ii}^{\ C\,V}$), presence of in-active implanted species ($C_{ii}^{\ impl}$) and one more stabilizing term ($D_{11}^{\ redistribution}$). But the value of $C_{11}^{tot}$ is still slightly negative except 2-4 keV ion energy which indicates the slow growth rate of parallel ripple pattern. The observed negligible negative value of $C_{11}^{tot}$ for energy 1.3 to 10 keV predicts no ripple formation in typical fluence up to $10^{18}$ ions/ cm$^2$. They also reported no well-defined ripple pattern for 3-10 keV mass selected Ar$^+$ ion beam. However, they found negative value of $C_{11}^{tot}$ for energy less than 1.3 keV and experimentally observed ripple pattern when the experiments were performed by an unanalyzed Ar beam from a microwave plasma ion source.

The pure mass selected Ar$^+$ ions could not generate instability in the low energy regime whereas in the unanalyzed beam, the presence of contaminations changes the surface chemistry by reacting with Si atoms. The in-equal sputtering yield of pure Si and compound Si generate additional instability. Similarly, ion-induced mass redistribution is also altered and increases the surface instability. It was observed that the Si$_3$N$_4$ compound formed during 12 keV N$^+$ bombardments on Si and unequal sputtering of Si$_3$N$_4$ and Si generated the surface instability [39]. We recently have reported the ripple pattern formation due to preferential sputtering of different elements present on multi-elemental mica by 12 keV mass analyzed Ar bombardment [60]. But, for Si like mono-elemental surface, low energy noble ions could not generate sufficient instability to form a pattern. Hence only a flat surface with very low roughness is formed even after long time bombardment. It was previously reported that reactive O or C or N ion bombardment could generate instability on Si surface for ripple pattern formation[18,19,39,61-64]. That instability comes from reactive ion induced chemical compound formation with Si. But Ar being inert could not form any chemical compound with Si upto 20 keV ion bombardments. The ion beam used for very low energy (~ 500 eV) Ar bombardment is generally mass unanalyzed. Therefore, observed ripple pattern on Si surface may be due to the additional instability originated from the beam impurities. It was observed the SiC formation by 1-2 keV mass unanalyzed ion bombardment [31]. So, the chemical nature of the surface patterns formed by low energy ion beam must be studied for the conclusion. The present study on the effect of pure and contaminated $Ar^+$ ion beam bombardment on Si establishes the fact that the $Ar^+$ ion beam (up to 10 keV) induced Si pattern formation is mainly due to the reactive contaminations. On the other way, we can conclude that low energy ion bombardment could generate pattern formation on Si only when the chemical phase of Si is changed like the case of N,O and C ion bombardment.

**Conclusion**

In summary, we have experimentally observed the mass analyzed and without mass analyzed $Ar^+$ ion beam induced pattern formation on Si. For the mass analyzed ion bombardment, no chemical change of the Si surface is taken place and a flat surface is formed whereas, for the mass unanalyzed ion bombardment, surface chemistry is significantly changed and the well periodic ripple pattern is observed on the Si surface. This experimental understanding establishes the fact that the presence of contaminations with ion beam play the major role in the low energy $Ar^+$ ion induced ripple pattern formation on Si. Our results will give a boost to study the beam purity and the surface chemistry of the patterns formed by mass unanalyzed ion beam. It will also stimulate to improve the theoretical understanding by incorporating the chemical effect in the continuum model of low energy ion beam induced surface pattern formation.

## Acknowledgments

The authors are grateful to VECC Kolkata, DAE, Govt. of India for providing financial support to carry out the research. We thank Mr. Chinmay Giri, Mr. Sayed Masum and Sanket Haque for their help during ion irradiation. The authors also acknowledge Mr. Goutam Sarkar for XPS measurements.